\documentclass[aps,prb,reprint,amsmath,amssymb,superscriptaddress,footinbib,longbibliography]{revtex4-1}
\usepackage{graphicx}
\usepackage{textcomp} 
\usepackage{gensymb}
\usepackage[utf8]{inputenc}
\usepackage{xcolor}

\begin{document}

\title{Everlasting bubbles and liquid films resisting drainage, evaporation and nuclei-induced bursting.}

\author{Aymeric Roux}
\affiliation{Univ. Lille, CNRS, Centrale  Lille, Univ. Polytechnique Hauts-de-France, UMR 8520, IEMN, F59000 Lille, 
France}%
\author{Alexis Duchesne}
\affiliation{Univ. Lille, CNRS, Centrale  Lille, Univ. Polytechnique Hauts-de-France, UMR 8520, IEMN, F59000 Lille, 
France}%
\author{Michael Baudoin}
\email{Corresponding author: \mbox{michael.baudoin@univ-lille.fr}}
\affiliation{Univ. Lille, CNRS, Centrale  Lille, Univ. Polytechnique Hauts-de-France, UMR 8520, IEMN, F59000 Lille, 
France}%
\affiliation{Institut Universitaire de France, 1 rue Descartes, 75005 Paris}%

\begin{abstract}
Soap bubbles are by essence fragile and ephemeral. Depending on their composition and environment, bubble bursting can be triggered by gravity-induced drainage and/or the evaporation of the liquid and/or the presence of nuclei. They can also shrink due to the diffusion of the inner gas in the outside atmosphere induced by Laplace overpressure. In this paper, we design bubbles made of a composite liquid film able to neutralize all these effects and keep their integrity for more than one year in a standard atmosphere. The unique properties of this composite film are rationalized with a nonlinear model and used to design complex objects.\end{abstract}

\maketitle

\onecolumngrid

\begin{figure}[htbp]
\centering
    \includegraphics[width=0.65\textwidth]{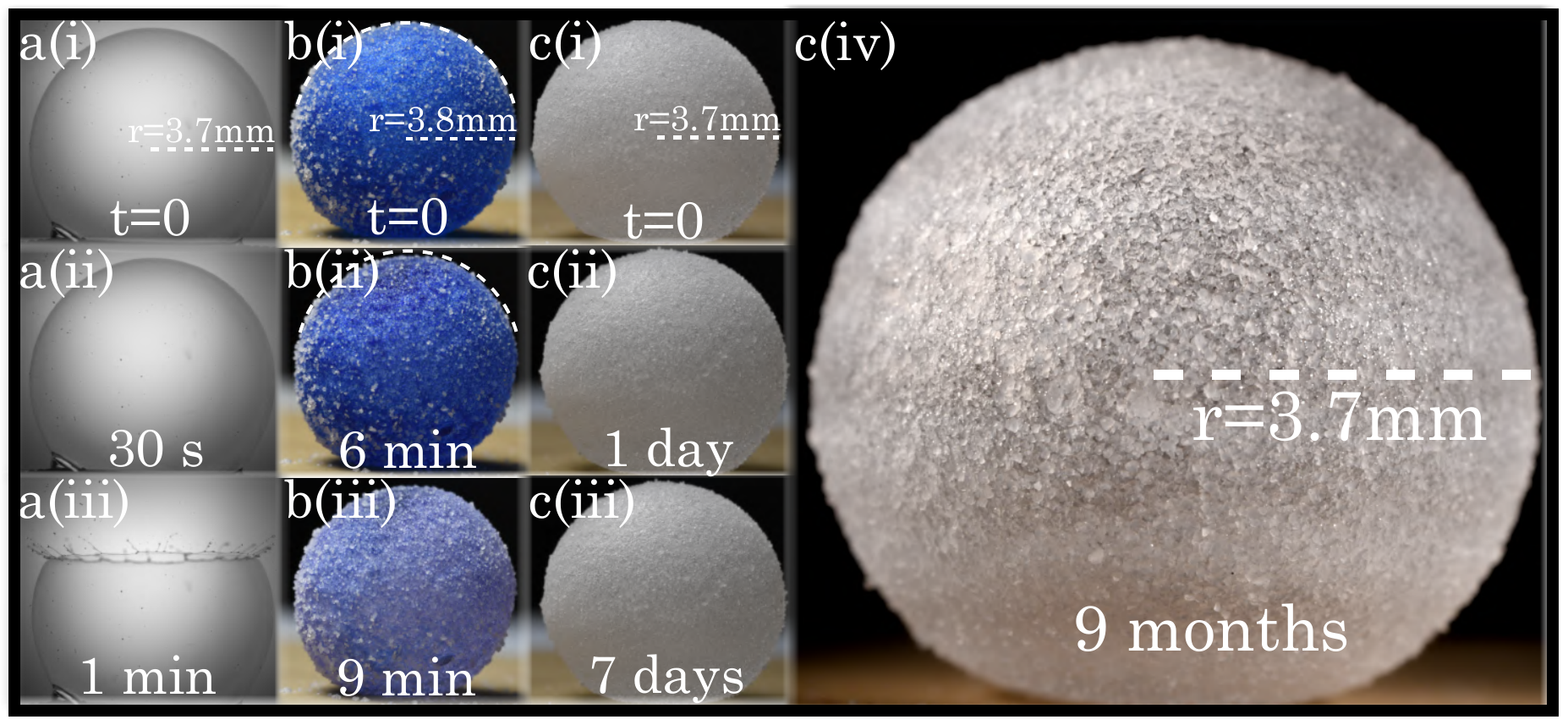}
    \caption{Comparison of the lifetimes of three types of bubbles: a) Soap bubble bursting after 1 min. b) Water gas marble covered by partially-wetting micro-particles, whose shell rupture occurs after 6 min and complete drying after 9 min.  Dashed line is used as a guide to visualize bubble opening and blue dye to visualize liquid drying. c) Water/glycerol gas marble perfectly intact in a standard atmosphere after $\sim9$ months (285 days). This bubble kept its integrity for more than one year ($465$ days) \cite{note2}. As can be seen on Fig.\ref{figSMAT1}, the shell of a water gas marble once dried collapses like a sand pile (Movie M1), while the shell of the water/glycerol gas marble is still liquid and spherical  and reacts as a liquid film when punctured (Movie M2).}
\label{fig1}
\end{figure}

\twocolumngrid

Under a standard atmosphere, soap bubbles typically burst in minutes (Fig.\ref{fig1}a) due to the effect of gravity-induced drainage and/or liquid evaporation and/or nucleï induced inception depending on their composition. "Bare" viscous bubbles bursting follows gravity-induced drainage of the liquid, resulting in shell thinning down to a limit thickness of the order of tens of nanometers, wherein spontaneous breaking occurs \cite{bare_bubbles}. The sliding condition of the liquid on air leads to plug flow in the shell and hence characteristic lifetime $\tau = \mu / \rho_l g R $, which does not rely on the liquid shell thickness $e$ but rather on the radius of curvature of the bubble $R$, with $\rho_l$ the liquid density, $g$ the gravitational acceleration and $\mu$ the liquid dynamic viscosity. The addition of surfactants to produce soap bubbles can lead to completely different pictures depending on the concentration of surfactants \cite{p_mysels_1959,modeling_soap}: At intermediate surfactant concentrations, bubble ageing results from a complex interplay between gravity and capillary induced drainage, Marangoni stresses due to gradients of surfactant concentration and bubble evaporation. At large surfactant concentration, the surfactant rigidifies the film surface leading to no-slip boundary condition, Poiseuille flow and hence considerably slowed down drainage. Nevertheless, while surfactant drastically increase the lifetime of soap bubbles, evaporation and/or the presence of nuclei will  eventually lead to their rupture.  Hence long bubbles lifetimes are only reported in carefully controlled atmosphere \cite{s_grosse_1969}, wherein dusts are suppressed, the atmosphere is saturated to prevent evaporation, and the level of mechanical vibration is controlled. And even in these conditions, bubbles shrinkage resulting from the diffusion of the inner air in the outside atmosphere due to Laplace overpressure still occurs \cite{s_grosse_1969}.

\begin{figure}[htbp]
\centering
    \includegraphics[width=0.45\textwidth]{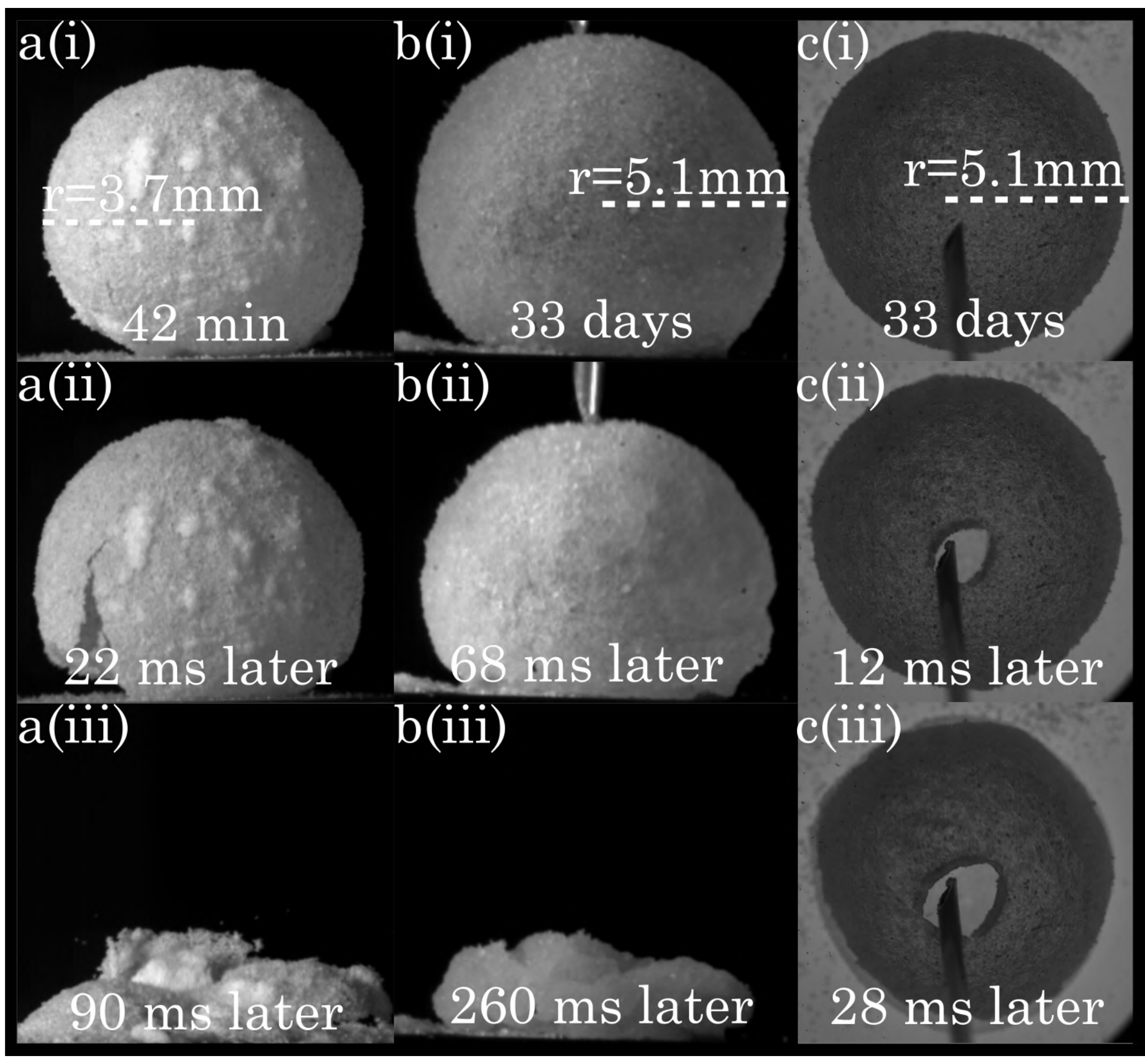}
    \caption{(a) Lateral view of a water gas marble naturally collapsing after a few tens of minutes. (b) Top view  and (c) lateral view  of two one-month old water/glycerol gas marbles (initial glycerol mass ratio $\omega_{g0} = 0.60$) punctured with a needle. \label{figSMAT1}}
\end{figure}

 \begin{figure*}[htbp]
\centering
    \includegraphics[width=0.7\textwidth]{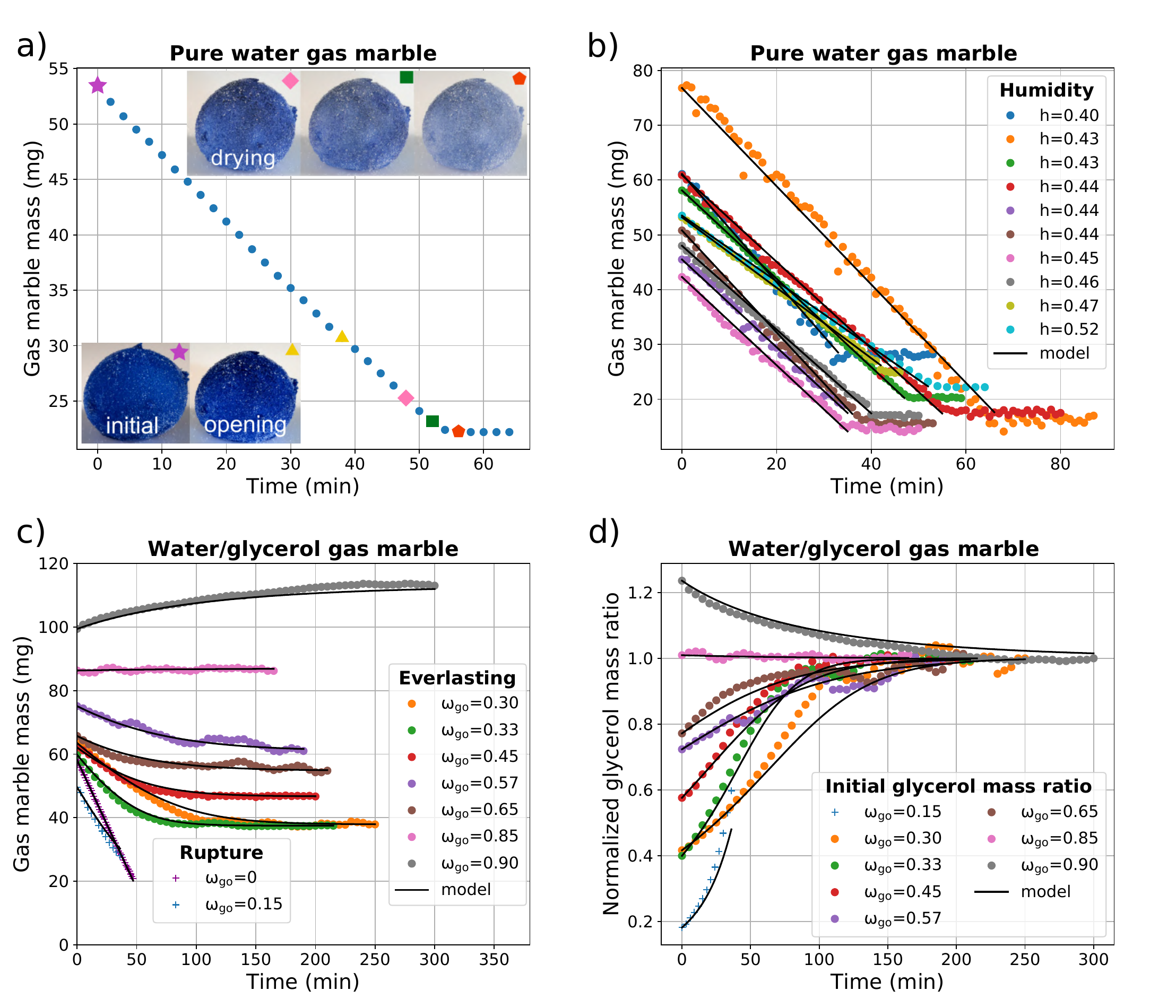}
    \caption{Evolution of gas marbles in a standard atmosphere (temperature $T =21.0\pm0.5^\degree C$). a) Evolution of the mass of a single water gas marble as a function of time at ambient humidity $h = 0.52\pm0.01$. Drying is reflected by the change in the bubble color and bubble opening (yellow triangle on the curve) by the brutal change in the bubble shape (Movie M4). The plateau region (after 54 min) occurs when the water is entirely evaporated. b) Evolution of the mass of pure water gas marbles as a function of time at different relative humidity. c)  Evolution of the mass of water/glycerol gas marbles as a function of time for different initial glycerol mass ratio $\omega_{go}$. d) Normalized glycerol mass ratio ($\omega_g/\omega_{\mathrm{geq}}$) content of gas marble as a function of time (same experimental data as in Fig.\ref{fig2}c). Black curves: evaporation/absorption model introduced in this paper.}
\label{fig2}
\end{figure*}

Adding partially-wetting particles to liquid/air interfaces \cite{rev_binks} has been investigated as a mean to strengthen their resistance to mechanical stresses providing a material with properties at the crossroad between solids and liquids \cite{elasticity_raft}. When drops and bubbles are covered with such a composite interface, some extraordinary properties have been reported such as: non-stick droplets  \cite{liquid_marbles,rpp_quere_2005,sm_mchale_2011} able to roll on solid surfaces, "armoured bubbles"\cite{anomalous_capillary} in a liquid  stable to dissolution \cite{dissolution_arrest} and which can sustain metastable \cite{sm_prabhudesai_2017} non-spherical shapes \cite{non-sperical_bubble,capillary_tailoring} or "gas marbles" (air bubbles) \cite{much_stronger,low_permeabiblity} supporting positive and negative relative pressure one order of magnitude larger than the Laplace pressure. The addition of particles to an interface can also lead to some counter-intuitive behaviours such as surface energy driven fingering instability in reversed Saffman-Taylor configuration \cite{prl_bihi_2016} or films growing ahead of a liquid meniscus moving in a capillary tube \cite{capillary_tailoring}. 

Nevertheless, (i) the effect of such a composite interface on the lifetime of air bubbles has not been investigated and (ii) the single addition of particles to a bubble shell cannot prevent its evaporation (Fig.\ref{fig1}b) and hence its bursting. In this paper \cite{arxiv_roux_2021}, we show that covering a bubble water shell with micro-particles inhibit gravity-induced drainage and that further adding glycerol leads to a stable state, wherein the evaporation of water is counterbalanced by the hygroscopicity of glycerol, which absorbs water molecules contained in the ambient air. This results in bubbles (Fig.\ref{fig1}c)  which can keep their integrity in a standard atmosphere for more than one year, with no significant evolution of their radius \cite{note2}. This behaviour is rationalized with a nonlinear model able to reproduce quantitatively the evolution of the mass of the bubble and predict their fate depending on the initial composition of the bubble and the surrounding atmosphere.

\noindent \textbf{Neutralizing drainage.} "Gas marbles" are by definition gas bubbles whose composite shell is made of liquid and partially-wetting particles \cite{much_stronger,low_permeabiblity}. Here, we first study the mass evolution and lifetime of water gas marbles for different ambient humidities (Fig.\ref{fig2}a and \ref{fig2}b). Gas marbles of radius $5.1 \pm 0.1$mm are produced with the simple following process (movie M3): First, some partially-wetting particles (polyamide-11 particles, average radius $r_p = 80\pm20 \; \mu$m, contact angle with water $71 \pm 3 \degree$\cite{capillary_tailoring}),  are spread at the surface of a water bath to form a jammed granular raft. Second, a controlled volume of air is injected with a syringe below the granular raft, leading to the formation of air bubbles whose upper interface is covered with particle. Third, the bubble is pushed with a spoon toward the surrounding particle raft and made roll over it to cover its whole surface. Finally, the bubble is extracted and placed over a hydrophobic (Teflon) or superhydrophobic (nanopillar) substrate. The evolution of the gas marble mass and shape are respectively monitored  with a FAS224 Fisherbrand high precision balance and a Nikon D850 camera with a 25 mm ultra macro Laowa lens. The relative humidity is measured with a A+life TH1818 hygrometer. Fig.\ref{fig2}a shows the evolution of the mass of a single water gas marble. Bubble drying is visualized with the addition of a blue dye and associated color change, while bubble opening (marked with a yellow triangle on Fig.\ref{fig2}a) can be seen by the brutal change in the bubble shape when the bubble shell is no more airtight (Movie M4). These data show (i) that the bubble mass decreases linearly until its complete drying (plateau region after 54 minutes on Fig.\ref{fig2}a) and (ii) that the gas marble opening (yellow triangle) occurs a few minutes before its complete drying, thus underlying that the water gas marble lifetime is mainly determined by the evaporation process. This scenario is further confirmed by numerous experiments performed for different ambient humidities (Fig.\ref{fig2}b). Hence, the presence of the particle shell inhibits the drainage of the liquid, which, in absence of the particles, would provoke bubble bursting after a characteristic time of the order of $\tau =  \mu / \rho_l g R \sim 20 \, \mu$s \cite{bare_bubbles}, with $R$ the bubble radius, $\rho_l$ the liquid density, $g$ the gravitational acceleration and $\mu$ the liquid viscosity. Note that the bubble lifetime variability on Fig.\ref{fig2}b at same ambient humidity is due to the fact that the initial film thickness is not controlled during the gas marble formation process. 
\begin{figure}[htbp]
\centering
    \includegraphics[width=0.4\textwidth]{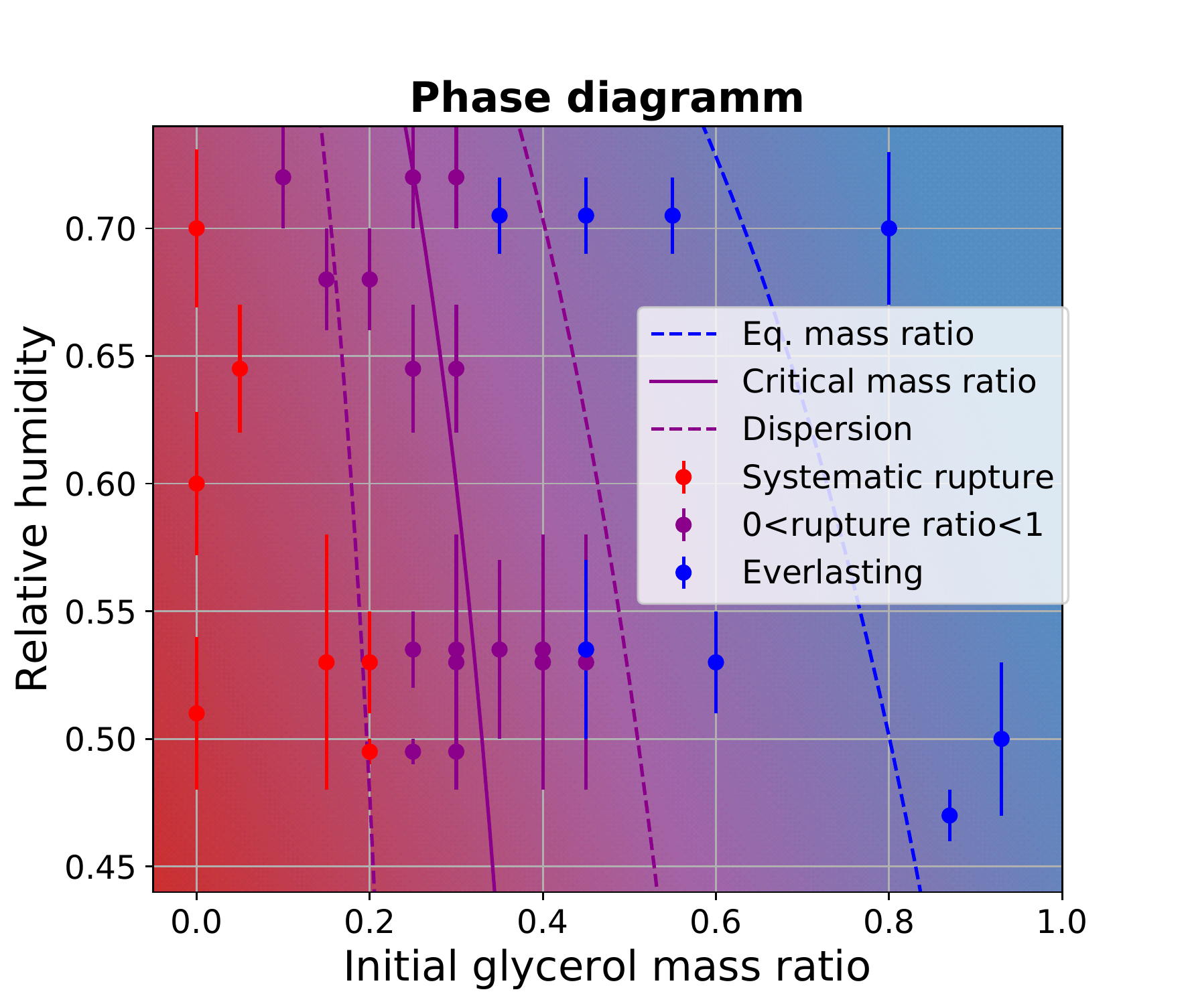}
    \caption{Diagram of gas marble fate after six hours as a function of the initial glycerol mass ratio and the relative humidity. Each point corresponds to statistics made on at least ten gas marbles. Red: Systematic rupture of the bubble. Purple: Part of the bubble sample rupture. Blue: All gas marble keeping their integrity after six hours. The vertical bar indicates the variation of the relative humidity for each set of experiment. The dashed blue curve entitled "Eq. mass ratio" represents the equilibrium glycerol mass ratio $\omega_{g\mathrm{eq}}(h)$ and the purple continuous line the critical glycerol mass ratio $\omega_{gco}$.}
\label{fig3}
\end{figure}

\noindent \textbf{Neutralizing evaporation.} Designing "everlasting" bubbles, i.e. bubbles with very long lifetimes, requires to further neutralize evaporation. This problem is solved by adding  glycerol to water. Indeed, glycerol is a liquid with a high concentration of hydroxyl groups, which have a strong affinity with water molecules with whom they create hydrogen bonds \cite{jml_chen_2009}. This mechanism is at the origin of glycerol's hygroscopicity (ability to absorb water molecules contained in air), which can compensate for water evaporation \cite{glycerol_marbles}. To test this hypothesis, bubbles were synthesized with the same process as described in the previous section but with a water/glycerol mixture. The evolution of the mass of these gas marbles for different initial glycerol mass ratio $\omega_{go}$ and different values of the ambient relative humidity $h$ (given in SI) is shown on Fig.\ref{fig2}c. This figure shows that the shell of water/glycerol gas marbles with the lowest initial glycerol concentrations ($\omega_{go} \in [0 , 0.15]$) rupture after lifetimes of the order of 50 min, while the bubbles with higher glycerol concentration tend toward a steady state either by initially losing water ($\omega_{go} \in [0.30 , 0.65]$), keeping a mass relatively constant ($\omega_{go} = 0.85$) or absorbing water from the humidity contained in the air ($\omega_{go} = 0.9$). Since glycerol is a non-volatile liquid, the mass variations of these gas marbles are solely due to water evaporation or absorption. The evolution toward a steady state with an equilibrium between the water and glycerol content is even clearer when we study the evolution of the normalized glycerol mass ratio (Fig.\ref{fig2}d), which tends toward a constant value (slight fluctuations are due to recorded fluctuations in the air humidity). We further studied the fate of gas marbles as a function of the initial glycerol mass ratio and relative humidity (Fig.\ref{fig3}). When a water/glycerol mixture is in contact with humid air with a relative humidity $h$, it will tend toward an equilibrium glycerol mass ratio $\omega_{geq}$ materialized by the blue dotted line on Fig.\ref{fig3}. The diagram shows that when the initial glycerol mass ratio $\omega_{go}$ is close to this equilibrium value, the gas marble evolves toward a stable equilibrium state (everlasting bubbles) either by losing some liquid (when $\omega_{go} < \omega_{geq}$) or absorbing some liquid (when $\omega_{go} > \omega_{geq}$) \textcolor{blue}{\cite{note}}. 
When the initial glycerol mass ratio is too far from this equilibrium value, the gas marble systematically ruptures.

\begin{figure*}[htbp]
\centering
    \includegraphics[width=0.7\textwidth]{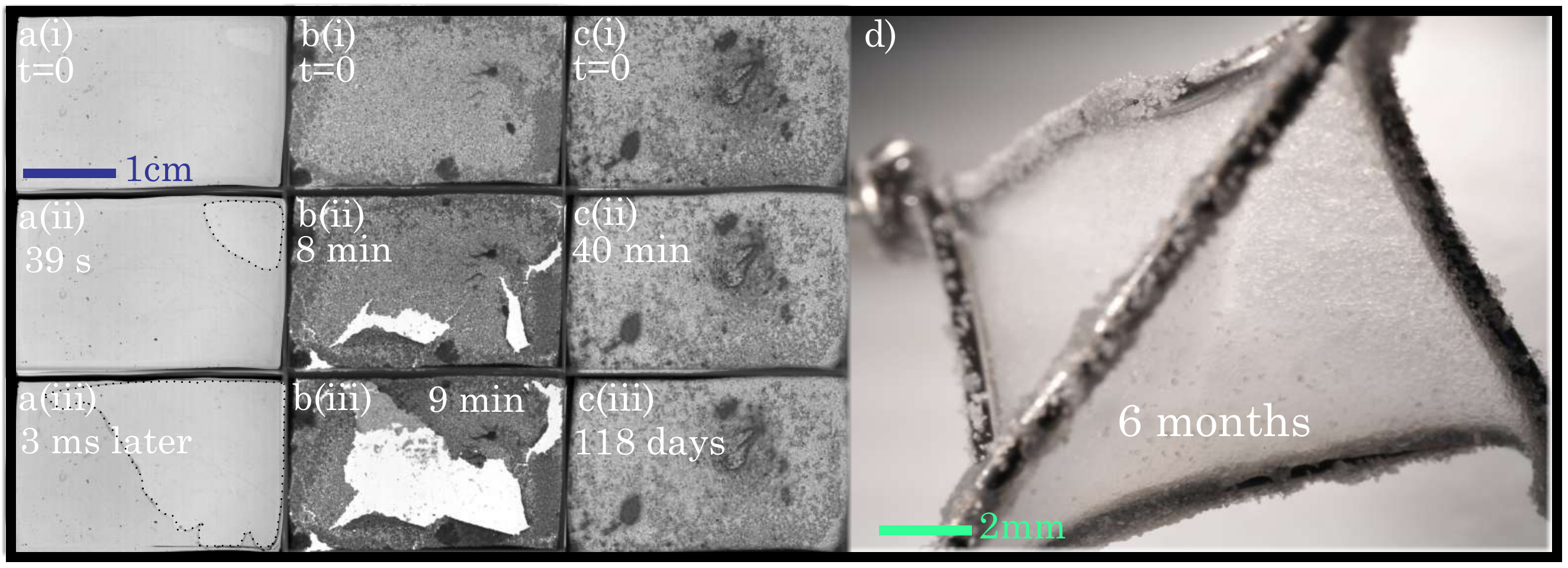}
    \caption{Lifetime of liquid films attached to a metallic frame (a,b,c: horizontal frame. d. pyramidal frame) a) Soap film made of water and surfactant. b) Water film covered with partially wetting particles. c) Water/glycerol liquid film (initial glycerol mass concentration $\omega_{go} = 0.80$) covered with particles. d) 3D object created with a pyramidal frame supporting a water/glycerol liquid film covered with particles and perfectly intact after 203 days ($\sim 6$ months). When these lines are written the pyramidal film is still perfectly intact in our laboratory 378 days after its formation.}
\label{fig4}
\end{figure*}

\noindent \textbf{Model.} To rationalise this behaviour, we introduce a model able to quantitatively reproduce the experimental trends described in Fig.\ref{fig2} and \ref{fig3} without any adjustable parameter. Since glycerol is a non-volatile liquid, the liquid mass $m_l$ evolution of a water/glycerol gas marble is set by an equilibrium between the flux of water evaporation and the flux of water absorption due to the hygroscopicity of glycerol: $dm_l / dt = \phi_{\mathrm{abs}} - \phi_{\mathrm{evap}}.$ Water evaporation flux is driven by the vapor concentration gradient and hence is proportional to $(1-h)$, with $h$ the relative humidity of air far from the bubble \cite{convection_evaporating,big_bubbles}. Furthermore, since the shell liquid is a mixture of glycerol and water, the water evaporation flux will be proportional to the quantity of water molecules present at the bubble surface. Since the diffusion time $\tau_d = e^2 / D_{wg} \lesssim 3$ min (with $e \sim 150\, \mu$m  the thickness of the film estimated from the surface of the bubble and its mass and $D_{wg} \gtrsim 10^{-10}$ m$^2$s$^{-1}$ \cite{diffusion_coefficient}  the diffusion coefficient of water molecules into the water/glycerol mixture)  is short compared to the characteristic time $\tau_c \gtrsim 100$ min required for a bubble to reach its equilibrium state, the water/glycerol mixture in the film can be supposed to be homogeneous at the bubble mass evolution timescale. Hence, the water evaporation rate will be proportional to the concentration of water in the water/glycerol mixture $(1-\omega_g)$ with $\omega_g$ the glycerol mass ratio. We thus obtain: $\phi_{\mathrm{evap}} = k_e (1-h) (1-\omega_g),$ with $k_e$ a constant to be determined. Conversely the absorption of water through the formation of hydrogen bonds with glycerol molecules will be proportional to the concentration of water in the air $h$ and to the concentration of glycerol in the liquid film $\omega_g$, so that $\phi_{\mathrm{abs}} = k_a h \omega_g.$ The final equation: $dm_l/dt = k_a h \omega_g - k_e (1-h) (1-\omega_g)$ relies on two unknown coefficients $k_e$ and $k_a$, which depend on the evaporation regime, absorption efficiency and bubble geometry. The first coefficient can be determined by examining the case of pure water gas marble. In this case $\phi_{\mathrm{evap}} = k_e (1-h)$. A first validation of this model is that the linear trend is well recovered on Fig.\ref{fig2}b. The value of the coefficient $k_e = 2.3 \pm 0.2 \times 10^{-8}$ kg s$^{-1}$ is obtained by taking the median value of the slopes in Fig.\ref{fig2}b, which enables to obtain good fits for all the curves. Note that this coefficient is not a universal parameter and depends on the complex physics behind the evaporation process and the bubble geometry. The second coefficient is determined by examining the equilibrium state $dm_l/dt = 0$. In this case, the mass concentration of glycerol in the water/glycerol mixture reaches a well established equilibrium value \cite{properties_glycerine}, which depends on the relative humidity of the ambient air. Our model predicts a relation between the equilibrium glycerol mass ratio $\omega_{geq}$ and the relative humidity given by: $\omega_{geq} = [1 + k_a h/ k_e (1-h)]^{-1}.$ A second validation of this model is that this law fits well the abacus available in the literature \cite{properties_glycerine} (Fig.2 in SI), with $k_a/k_e = 0.248$, giving good confidence in the dependence of the evaporation and absorption fluxes on the glycerol concentration $\omega_g$ and relative humidity $h$. Now, that the two coefficients $k_a$ and $k_e$, are determined we can compare the model predictions to the experimentally measured gas marble mass evolution. This model recovers quantitatively the trends observed on Fig.\ref{fig2}c and \ref{fig2}d. From this model, we can also derive a criterion for the transition between ephemeral and everlasting bubbles (Fig.\ref{fig3}) depending on the initial glycerol concentration $\omega_{go}$ and relative humidity $h$. Assuming that bubbles rupture occurs when the bubble shell liquid mass reaches a critical value $m_{lc}$, a bubble will be everlasting if its equilibrium liquid mass $m_{leq}$ is larger than $m_{lc}$ and ephemeral if $m_{leq}$ is smaller than $m_{lc}$, with a transition between these regimes for $m_{leq} = m_{lc}$. Since the mass of glycerol $m_g$ is constant, we have $m_g = m_{leq} \omega_{geq} = m_{lo} \omega_{go}$. Hence for $m_{leq} = m_{lc}$, we obtain $\omega_{goc} = m_{lc} \omega_{geq}(h) / m_{lo} $ with $\omega_{geq}(h)$ determined previously. The critical liquid mass $m_{lc} = 14\pm4$ mg is determined by measuring the bubble total mass $m_c = 32\pm3$ when its rupture occurs (measured here for 4 different gas marbles) and subtracting the particle mass $m_p = 18\pm2$ mg (value extracted from Fig.\ref{fig2}b after total evaporation of water). Since we cannot control the liquid content during the gas marble formation process, an ephemeral gas marble initial content $m_{lo}$ varies typically between $22$ mg and $57$ mg. The plot of the equation $\omega_{goc} = m_{lc} \omega_{geq}(h) / m_{lo} $  on Fig.\ref{fig3}, shows that this model gives a good criterion for the transition between ephemeral and everlasting bubble (the continuous purple line corresponds to the average value $m_{lo} = 34$ mg, while the purple dashed lines correspond to $m_{lo} = 22$mg and $57$mg respectively and account for the dispersion of the initial state).

\noindent \textbf{Perspectives: composite films.} In this paper, we have shown that air bubbles keeping their integrity for more than one year can be produced in a simple way by replacing surfactants by partially-wetting particles and water by a water/glycerol mixture. Indeed, partially-wetting particles prevent drainage of the liquid due to gravity, while the hygroscopicity of glycerol-water mixtures counterbalances evaporation, which are the two main mechanisms at the origin of classical bubble bursting. In addition, (i) the neutralization of these two phenomena and the presence of the particle shell enables to make bubble insensitive to the nuclei contained in the air and (ii) no evolution of the bubble radius over the bubble life was observed, indicating that bubble shrinkage due to Laplace overpressure is also suppressed. Beyond bubbles, this work unveils a robust composite liquid film, which can be used to create a wealth of different objects. Indeed, when a metallic frame is dipped below a liquid surface covered with a layer of jammed particles and lifted slowly up to the surface, it captures the particle-covered film (see Fig.\ref{fig4}). While marginal pinching due to the presence of the frame and surfactants leads to soap films bursting near the edge after a few seconds (Fig.\ref{fig4}.a), the replacement of surfactants by partially-wetting particles leads to the apparition of evaporation induced cracks after several minutes only (Fig.\ref{fig4}.b), and further addition of glycerol leads to long lasting liquid film stable for at least  several  months (Fig.\ref{fig4}.c). Following this process, complex objects such as everlasting pyramidal films with lifetimes of more than one year (378 days when these lines are written) can easily be manufactured (Fig.\ref{fig4}.d) paving the way toward the design of new fluidic objects with unexplored physical and chemical properties. This work would be ideally completed with a study on a large set of parameters of the effect of the composition of the shell and of the ambient atmosphere on the bubble stability. Finally, a puzzling question remains: does the Laplace overpressure inside the bubble vanishes over time and if it does, how does the bubble maintains its shape and size? We are currently trying to provide a convincing answer to this question.

\noindent \textbf{Acknowledgements:}
We would like to thank F. Zoueshtiagh for useful discussions and technical assistance. This work was supported by ISITE-ULNE (ERC Generator program), Institut Universitaire de France and Renatech network. \\


\begin{thebibliography}{0}%
\makeatletter
\providecommand \@ifxundefined [1]{%
 \@ifx{#1\undefined}
}%
\providecommand \@ifnum [1]{%
 \ifnum #1\expandafter \@firstoftwo
 \else \expandafter \@secondoftwo
 \fi
}%
\providecommand \@ifx [1]{%
 \ifx #1\expandafter \@firstoftwo
 \else \expandafter \@secondoftwo
 \fi
}%
\providecommand \natexlab [1]{#1}%
\providecommand \enquote  [1]{``#1''}%
\providecommand \bibnamefont  [1]{#1}%
\providecommand \bibfnamefont [1]{#1}%
\providecommand \citenamefont [1]{#1}%
\providecommand \href@noop [0]{\@secondoftwo}%
\providecommand \href [0]{\begingroup \@sanitize@url \@href}%
\providecommand \@href[1]{\@@startlink{#1}\@@href}%
\providecommand \@@href[1]{\endgroup#1\@@endlink}%
\providecommand \@sanitize@url [0]{\catcode `\\12\catcode `\$12\catcode
  `\&12\catcode `\#12\catcode `\^12\catcode `\_12\catcode `\%12\relax}%
\providecommand \@@startlink[1]{}%
\providecommand \@@endlink[0]{}%
\providecommand \url  [0]{\begingroup\@sanitize@url \@url }%
\providecommand \@url [1]{\endgroup\@href {#1}{\urlprefix }}%
\providecommand \urlprefix  [0]{URL }%
\providecommand \Eprint [0]{\href }%
\providecommand \doibase [0]{http://dx.doi.org/}%
\providecommand \selectlanguage [0]{\@gobble}%
\providecommand \bibinfo  [0]{\@secondoftwo}%
\providecommand \bibfield  [0]{\@secondoftwo}%
\providecommand \translation [1]{[#1]}%
\providecommand \BibitemOpen [0]{}%
\providecommand \bibitemStop [0]{}%
\providecommand \bibitemNoStop [0]{.\EOS\space}%
\providecommand \EOS [0]{\spacefactor3000\relax}%
\providecommand \BibitemShut  [1]{\csname bibitem#1\endcsname}%
\let\auto@bib@innerbib\@empty
\end{thebibliography}%


\begin{thebibliography}{26}%
\makeatletter
\providecommand \@ifxundefined [1]{%
 \@ifx{#1\undefined}
}%
\providecommand \@ifnum [1]{%
 \ifnum #1\expandafter \@firstoftwo
 \else \expandafter \@secondoftwo
 \fi
}%
\providecommand \@ifx [1]{%
 \ifx #1\expandafter \@firstoftwo
 \else \expandafter \@secondoftwo
 \fi
}%
\providecommand \natexlab [1]{#1}%
\providecommand \enquote  [1]{``#1''}%
\providecommand \bibnamefont  [1]{#1}%
\providecommand \bibfnamefont [1]{#1}%
\providecommand \citenamefont [1]{#1}%
\providecommand \href@noop [0]{\@secondoftwo}%
\providecommand \href [0]{\begingroup \@sanitize@url \@href}%
\providecommand \@href[1]{\@@startlink{#1}\@@href}%
\providecommand \@@href[1]{\endgroup#1\@@endlink}%
\providecommand \@sanitize@url [0]{\catcode `\\12\catcode `\$12\catcode
  `\&12\catcode `\#12\catcode `\^12\catcode `\_12\catcode `\%12\relax}%
\providecommand \@@startlink[1]{}%
\providecommand \@@endlink[0]{}%
\providecommand \url  [0]{\begingroup\@sanitize@url \@url }%
\providecommand \@url [1]{\endgroup\@href {#1}{\urlprefix }}%
\providecommand \urlprefix  [0]{URL }%
\providecommand \Eprint [0]{\href }%
\providecommand \doibase [0]{http://dx.doi.org/}%
\providecommand \selectlanguage [0]{\@gobble}%
\providecommand \bibinfo  [0]{\@secondoftwo}%
\providecommand \bibfield  [0]{\@secondoftwo}%
\providecommand \translation [1]{[#1]}%
\providecommand \BibitemOpen [0]{}%
\providecommand \bibitemStop [0]{}%
\providecommand \bibitemNoStop [0]{.\EOS\space}%
\providecommand \EOS [0]{\spacefactor3000\relax}%
\providecommand \BibitemShut  [1]{\csname bibitem#1\endcsname}%
\let\auto@bib@innerbib\@empty
\bibitem [{not({\natexlab{a}})}]{note2}%
  \BibitemOpen
  \href@noop {} {} ({\natexlab{a}}),\ \bibinfo {note} {Note
  that our oldest bubble ruptured after 465 days. We believe that this rupture
  can be attributed to the development of "life" into our bubble since the
  bubbles became slightly green during the last month, which would not be
  surprising since (i) the bubble is made of water and glycerol which is a
  favorable environment for the development of fungi and bacteria and (ii) we
  did not take any precaution to avoid the pollution of the bubble with living
  organisms.}\BibitemShut {Stop}%
\bibitem [{\citenamefont {Debr{\'e}geas}\ \emph {et~al.}(1998)\citenamefont
  {Debr{\'e}geas}, \citenamefont {de~Gennes},\ and\ \citenamefont
  {Brochard-Wyart}}]{bare_bubbles}%
  \BibitemOpen
  \bibfield  {author} {\bibinfo {author} {\bibfnamefont {G.}~\bibnamefont
  {Debr{\'e}geas}}, \bibinfo {author} {\bibfnamefont {P.-G.}\ \bibnamefont
  {de~Gennes}}, \ and\ \bibinfo {author} {\bibfnamefont {F.}~\bibnamefont
  {Brochard-Wyart}},\ }\bibfield  {title} {\enquote {\bibinfo {title} {The life
  and death of "bare" viscous bubbles},}\ }\href {\doibase
  10.1126/science.279.5357.1704} {\bibfield  {journal} {\bibinfo  {journal}
  {Science}\ }\textbf {\bibinfo {volume} {279}},\ \bibinfo {pages} {1704--1707}
  (\bibinfo {year} {1998})}\BibitemShut {NoStop}%
\bibitem [{\citenamefont {Mysels}\ \emph {et~al.}(1950)\citenamefont {Mysels},
  \citenamefont {Shinoda},\ and\ \citenamefont {Frankel}}]{p_mysels_1959}%
  \BibitemOpen
  \bibfield  {author} {\bibinfo {author} {\bibfnamefont {K.J.}\ \bibnamefont
  {Mysels}}, \bibinfo {author} {\bibfnamefont {K.}~\bibnamefont {Shinoda}}, \
  and\ \bibinfo {author} {\bibfnamefont {S.}~\bibnamefont {Frankel}},\
  }\href@noop {} {\emph {\bibinfo {title} {Soap films, Studies of their
  thinning and a bibliography}}}\ (\bibinfo  {publisher} {Pergamon},\ \bibinfo
  {year} {1950})\BibitemShut {NoStop}%
\bibitem [{\citenamefont {Schwartz}\ and\ \citenamefont
  {Roy}(1999)}]{modeling_soap}%
  \BibitemOpen
  \bibfield  {author} {\bibinfo {author} {\bibfnamefont {L.W.}\ \bibnamefont
  {Schwartz}}\ and\ \bibinfo {author} {\bibfnamefont {R.V.}\ \bibnamefont
  {Roy}},\ }\bibfield  {title} {\enquote {\bibinfo {title} {Modeling draining
  flow in mobile and immobile soap films},}\ }\href {\doibase
  10.1006/jcis.1999.6426} {\bibfield  {journal} {\bibinfo  {journal} {J.
  Colloid Interf. Sci.}\ }\textbf {\bibinfo {volume} {218}},\ \bibinfo {pages}
  {309--323} (\bibinfo {year} {1999})}\BibitemShut {NoStop}%
\bibitem [{\citenamefont {Grosse}(1969)}]{s_grosse_1969}%
  \BibitemOpen
  \bibfield  {author} {\bibinfo {author} {\bibfnamefont {A.V.}\ \bibnamefont
  {Grosse}},\ }\bibfield  {title} {\enquote {\bibinfo {title} {Soap bubbles:
  two years old and six centimeters in diameter},}\ }\href@noop {} {\bibfield
  {journal} {\bibinfo  {journal} {Science}\ }\textbf {\bibinfo {volume}
  {164}},\ \bibinfo {pages} {291--293} (\bibinfo {year} {1969})}\BibitemShut
  {NoStop}%
\bibitem [{\citenamefont {Binks}(2002)}]{rev_binks}%
  \BibitemOpen
  \bibfield  {author} {\bibinfo {author} {\bibfnamefont {B.P.}\ \bibnamefont
  {Binks}},\ }\bibfield  {title} {\enquote {\bibinfo {title} {Particles as
  surfactants - similarities and differences},}\ }\href {\doibase
  10.1016/S1359-0294(02)00008-0} {\bibfield  {journal} {\bibinfo  {journal}
  {Curr. Opin. Colloid In.}\ }\textbf {\bibinfo {volume} {7}},\ \bibinfo
  {pages} {21--41} (\bibinfo {year} {2002})}\BibitemShut {NoStop}%
\bibitem [{\citenamefont {Vella}\ \emph {et~al.}(2004)\citenamefont {Vella},
  \citenamefont {Aussillous},\ and\ \citenamefont
  {Mahadevan}}]{elasticity_raft}%
  \BibitemOpen
  \bibfield  {author} {\bibinfo {author} {\bibfnamefont {D.}~\bibnamefont
  {Vella}}, \bibinfo {author} {\bibfnamefont {P.}~\bibnamefont {Aussillous}}, \
  and\ \bibinfo {author} {\bibfnamefont {L.}~\bibnamefont {Mahadevan}},\
  }\bibfield  {title} {\enquote {\bibinfo {title} {Elasticity of an interfacial
  particle raft},}\ }\href {\doibase 10.1209/epl/i2004-10202-x} {\bibfield
  {journal} {\bibinfo  {journal} {Europhys. Lett.}\ }\textbf {\bibinfo {volume}
  {68}},\ \bibinfo {pages} {212--218} (\bibinfo {year} {2004})}\BibitemShut
  {NoStop}%
\bibitem [{\citenamefont {Aussillous}\ and\ \citenamefont
  {Quéré}(2001)}]{liquid_marbles}%
  \BibitemOpen
  \bibfield  {author} {\bibinfo {author} {\bibfnamefont {P.}~\bibnamefont
  {Aussillous}}\ and\ \bibinfo {author} {\bibfnamefont {D.}~\bibnamefont
  {Quéré}},\ }\bibfield  {title} {\enquote {\bibinfo {title} {Liquid
  marbles},}\ }\href {\doibase 10.1038/35082026} {\bibfield  {journal}
  {\bibinfo  {journal} {Nature}\ }\textbf {\bibinfo {volume} {411}},\ \bibinfo
  {pages} {924--927} (\bibinfo {year} {2001})}\BibitemShut {NoStop}%
\bibitem [{\citenamefont {Qu\'{e}r\'{e}}(2005)}]{rpp_quere_2005}%
  \BibitemOpen
  \bibfield  {author} {\bibinfo {author} {\bibfnamefont {D.}~\bibnamefont
  {Qu\'{e}r\'{e}}},\ }\bibfield  {title} {\enquote {\bibinfo {title}
  {Non-sticking drops},}\ }\href@noop {} {\bibfield  {journal} {\bibinfo
  {journal} {Rep. Prog. Phys.}\ }\textbf {\bibinfo {volume} {68}},\ \bibinfo
  {pages} {2495} (\bibinfo {year} {2005})}\BibitemShut {NoStop}%
\bibitem [{\citenamefont {McHale}\ and\ \citenamefont
  {Newton}(2011)}]{sm_mchale_2011}%
  \BibitemOpen
  \bibfield  {author} {\bibinfo {author} {\bibfnamefont {G.}~\bibnamefont
  {McHale}}\ and\ \bibinfo {author} {\bibfnamefont {M.I.}\ \bibnamefont
  {Newton}},\ }\bibfield  {title} {\enquote {\bibinfo {title} {Liquid marbles:
  principle and applications},}\ }\href@noop {} {\bibfield  {journal} {\bibinfo
   {journal} {Soft Matter}\ }\textbf {\bibinfo {volume} {7}},\ \bibinfo {pages}
  {5473--5481} (\bibinfo {year} {2011})}\BibitemShut {NoStop}%
\bibitem [{\citenamefont {Kam}\ and\ \citenamefont
  {Rossen}(1999)}]{anomalous_capillary}%
  \BibitemOpen
  \bibfield  {author} {\bibinfo {author} {\bibfnamefont {Seung~I}\ \bibnamefont
  {Kam}}\ and\ \bibinfo {author} {\bibfnamefont {William~R}\ \bibnamefont
  {Rossen}},\ }\bibfield  {title} {\enquote {\bibinfo {title} {Anomalous
  capillary pressure, stress, and stability of solids-coated bubbles},}\ }\href
  {\doibase https://doi.org/10.1006/jcis.1999.6107} {\bibfield  {journal}
  {\bibinfo  {journal} {J. Colloid Interf. Sci.}\ }\textbf {\bibinfo {volume}
  {213}},\ \bibinfo {pages} {329 -- 339} (\bibinfo {year} {1999})}\BibitemShut
  {NoStop}%
\bibitem [{\citenamefont {Abkarian}\ \emph {et~al.}(2007)\citenamefont
  {Abkarian}, \citenamefont {Subramaniam}, \citenamefont {Kim}, \citenamefont
  {Larsen}, \citenamefont {Yang},\ and\ \citenamefont
  {Stone}}]{dissolution_arrest}%
  \BibitemOpen
  \bibfield  {author} {\bibinfo {author} {\bibfnamefont {Manouk}\ \bibnamefont
  {Abkarian}}, \bibinfo {author} {\bibfnamefont {Anand~Bala}\ \bibnamefont
  {Subramaniam}}, \bibinfo {author} {\bibfnamefont {Shin-Hyun}\ \bibnamefont
  {Kim}}, \bibinfo {author} {\bibfnamefont {Ryan~J.}\ \bibnamefont {Larsen}},
  \bibinfo {author} {\bibfnamefont {Seung-Man}\ \bibnamefont {Yang}}, \ and\
  \bibinfo {author} {\bibfnamefont {Howard~A.}\ \bibnamefont {Stone}},\
  }\bibfield  {title} {\enquote {\bibinfo {title} {Dissolution arrest and
  stability of particle-covered bubbles},}\ }\href {\doibase
  10.1103/PhysRevLett.99.188301} {\bibfield  {journal} {\bibinfo  {journal}
  {Phys. Rev. Lett.}\ }\textbf {\bibinfo {volume} {99}},\ \bibinfo {pages}
  {188301} (\bibinfo {year} {2007})}\BibitemShut {NoStop}%
\bibitem [{\citenamefont {Prabhudesai}\ \emph {et~al.}(2017)\citenamefont
  {Prabhudesai}, \citenamefont {Bihi}, \citenamefont {Zoueshtiagh},
  \citenamefont {Jose},\ and\ \citenamefont {Baudoin}}]{sm_prabhudesai_2017}%
  \BibitemOpen
  \bibfield  {author} {\bibinfo {author} {\bibfnamefont {G.}~\bibnamefont
  {Prabhudesai}}, \bibinfo {author} {\bibfnamefont {I.}~\bibnamefont {Bihi}},
  \bibinfo {author} {\bibfnamefont {F.}~\bibnamefont {Zoueshtiagh}}, \bibinfo
  {author} {\bibfnamefont {J.}~\bibnamefont {Jose}}, \ and\ \bibinfo {author}
  {\bibfnamefont {M.}~\bibnamefont {Baudoin}},\ }\bibfield  {title} {\enquote
  {\bibinfo {title} {Nonspherical armoured bubble vibration},}\ }\href@noop {}
  {\bibfield  {journal} {\bibinfo  {journal} {Soft Matter}\ }\textbf {\bibinfo
  {volume} {13}},\ \bibinfo {pages} {3879} (\bibinfo {year}
  {2017})}\BibitemShut {NoStop}%
\bibitem [{\citenamefont {Subramaniam}\ \emph {et~al.}(2005)\citenamefont
  {Subramaniam}, \citenamefont {Abkarian}, \citenamefont {Mahadevan},\ and\
  \citenamefont {Stone}}]{non-sperical_bubble}%
  \BibitemOpen
  \bibfield  {author} {\bibinfo {author} {\bibfnamefont {A.B.}\ \bibnamefont
  {Subramaniam}}, \bibinfo {author} {\bibfnamefont {M.}~\bibnamefont
  {Abkarian}}, \bibinfo {author} {\bibfnamefont {L.}~\bibnamefont {Mahadevan}},
  \ and\ \bibinfo {author} {\bibfnamefont {H.A.}\ \bibnamefont {Stone}},\
  }\bibfield  {title} {\enquote {\bibinfo {title} {Non-spherical bubbles},}\
  }\href {\doibase 10.1038/438930a} {\bibfield  {journal} {\bibinfo  {journal}
  {Nature}\ }\textbf {\bibinfo {volume} {438}},\ \bibinfo {pages} {930}
  (\bibinfo {year} {2005})}\BibitemShut {NoStop}%
\bibitem [{\citenamefont {Zoueshtiagh}\ \emph {et~al.}(2014)\citenamefont
  {Zoueshtiagh}, \citenamefont {Baudoin},\ and\ \citenamefont
  {Guerrin}}]{capillary_tailoring}%
  \BibitemOpen
  \bibfield  {author} {\bibinfo {author} {\bibfnamefont {F.}~\bibnamefont
  {Zoueshtiagh}}, \bibinfo {author} {\bibfnamefont {M.}~\bibnamefont
  {Baudoin}}, \ and\ \bibinfo {author} {\bibfnamefont {D.}~\bibnamefont
  {Guerrin}},\ }\bibfield  {title} {\enquote {\bibinfo {title} {Capillary tube
  wetting induced by particles: Towards armoured bubbles tailoring},}\ }\href
  {\doibase 10.1039/c4sm01648c} {\bibfield  {journal} {\bibinfo  {journal}
  {Soft Matter}\ }\textbf {\bibinfo {volume} {10}},\ \bibinfo {pages}
  {9403--9412} (\bibinfo {year} {2014})}\BibitemShut {NoStop}%
\bibitem [{\citenamefont {Timounay}\ \emph
  {et~al.}(2017{\natexlab{a}})\citenamefont {Timounay}, \citenamefont
  {Pitois},\ and\ \citenamefont {Rouyer}}]{much_stronger}%
  \BibitemOpen
  \bibfield  {author} {\bibinfo {author} {\bibfnamefont {Yousra}\ \bibnamefont
  {Timounay}}, \bibinfo {author} {\bibfnamefont {Olivier}\ \bibnamefont
  {Pitois}}, \ and\ \bibinfo {author} {\bibfnamefont {Florence}\ \bibnamefont
  {Rouyer}},\ }\bibfield  {title} {\enquote {\bibinfo {title} {Gas marbles:
  Much stronger than liquid marbles},}\ }\href {\doibase
  10.1103/PhysRevLett.118.228001} {\bibfield  {journal} {\bibinfo  {journal}
  {Phys. Rev. Lett.}\ }\textbf {\bibinfo {volume} {118}},\ \bibinfo {pages}
  {228001} (\bibinfo {year} {2017}{\natexlab{a}})}\BibitemShut {NoStop}%
\bibitem [{\citenamefont {Timounay}\ \emph
  {et~al.}(2017{\natexlab{b}})\citenamefont {Timounay}, \citenamefont {Ou},
  \citenamefont {Lorenceau},\ and\ \citenamefont {Rouyer}}]{low_permeabiblity}%
  \BibitemOpen
  \bibfield  {author} {\bibinfo {author} {\bibfnamefont {Yousra}\ \bibnamefont
  {Timounay}}, \bibinfo {author} {\bibfnamefont {Even}\ \bibnamefont {Ou}},
  \bibinfo {author} {\bibfnamefont {Elise}\ \bibnamefont {Lorenceau}}, \ and\
  \bibinfo {author} {\bibfnamefont {Florence}\ \bibnamefont {Rouyer}},\
  }\bibfield  {title} {\enquote {\bibinfo {title} {Low gas permeability of
  particulate films slows down the aging of gas marbles},}\ }\href {\doibase
  10.1039/C7SM01444A} {\bibfield  {journal} {\bibinfo  {journal} {Soft Matter}\
  }\textbf {\bibinfo {volume} {13}},\ \bibinfo {pages} {7717--7720} (\bibinfo
  {year} {2017}{\natexlab{b}})}\BibitemShut {NoStop}%
\bibitem [{\citenamefont {Bihi}\ \emph {et~al.}(2016)\citenamefont {Bihi},
  \citenamefont {Baudoin}, \citenamefont {Faille},\ and\ \citenamefont
  {Zoueshtiagh}}]{prl_bihi_2016}%
  \BibitemOpen
  \bibfield  {author} {\bibinfo {author} {\bibfnamefont {I.}~\bibnamefont
  {Bihi}}, \bibinfo {author} {\bibfnamefont {J.E.}\ \bibnamefont {Baudoin},
  \bibfnamefont {M.~abd~Butler}}, \bibinfo {author} {\bibfnamefont
  {C.}~\bibnamefont {Faille}}, \ and\ \bibinfo {author} {\bibfnamefont
  {F.}~\bibnamefont {Zoueshtiagh}},\ }\bibfield  {title} {\enquote {\bibinfo
  {title} {Inverse {S}affman experiments with particles lead to capillary
  driven fingering instabilities},}\ }\href@noop {} {\bibfield  {journal}
  {\bibinfo  {journal} {Phys. Rev. Lett.}\ }\textbf {\bibinfo {volume} {117}},\
  \bibinfo {pages} {03450} (\bibinfo {year} {2016})}\BibitemShut {NoStop}%
\bibitem [{\citenamefont {Roux}\ \emph {et~al.}(2021)\citenamefont {Roux},
  \citenamefont {Duchesne},\ and\ \citenamefont {Baudoin}}]{arxiv_roux_2021}%
  \BibitemOpen
  \bibfield  {author} {\bibinfo {author} {\bibfnamefont {A.}~\bibnamefont
  {Roux}}, \bibinfo {author} {\bibfnamefont {A.}~\bibnamefont {Duchesne}}, \
  and\ \bibinfo {author} {\bibfnamefont {M.}~\bibnamefont {Baudoin}},\
  }\bibfield  {title} {\enquote {\bibinfo {title} {Bubbles and liquid films
  resisting drainage, evaporation and nuclei-induced bursting},}\ }\href@noop
  {} {\ ,\ \bibinfo {pages} {ArXiv:2103.15637} (\bibinfo {year}
  {2021})}\BibitemShut {NoStop}%
\bibitem [{\citenamefont {Chen}\ \emph {et~al.}(2009)\citenamefont {Chen},
  \citenamefont {Li}, \citenamefont {Song},\ and\ \citenamefont
  {Yang}}]{jml_chen_2009}%
  \BibitemOpen
  \bibfield  {author} {\bibinfo {author} {\bibfnamefont {C.}~\bibnamefont
  {Chen}}, \bibinfo {author} {\bibfnamefont {W.Z.}\ \bibnamefont {Li}},
  \bibinfo {author} {\bibfnamefont {Y.C.}\ \bibnamefont {Song}}, \ and\
  \bibinfo {author} {\bibfnamefont {J.}~\bibnamefont {Yang}},\ }\bibfield
  {title} {\enquote {\bibinfo {title} {Hydrogen bonding analysis of glycerol
  aqueous solutions: A molecular dynamics simulation study},}\ }\href@noop {}
  {\bibfield  {journal} {\bibinfo  {journal} {J Mol. Liq}\ }\textbf {\bibinfo
  {volume} {146}},\ \bibinfo {pages} {23--28} (\bibinfo {year}
  {2009})}\BibitemShut {NoStop}%
\bibitem [{\citenamefont {Lin}\ \emph {et~al.}(2019)\citenamefont {Lin},
  \citenamefont {Ma}, \citenamefont {Chen}, \citenamefont {Huang},
  \citenamefont {Wu},\ and\ \citenamefont {Takahara}}]{glycerol_marbles}%
  \BibitemOpen
  \bibfield  {author} {\bibinfo {author} {\bibfnamefont {X.}~\bibnamefont
  {Lin}}, \bibinfo {author} {\bibfnamefont {W.}~\bibnamefont {Ma}}, \bibinfo
  {author} {\bibfnamefont {L.}~\bibnamefont {Chen}}, \bibinfo {author}
  {\bibfnamefont {L.}~\bibnamefont {Huang}}, \bibinfo {author} {\bibfnamefont
  {H.}~\bibnamefont {Wu}}, \ and\ \bibinfo {author} {\bibfnamefont
  {A.}~\bibnamefont {Takahara}},\ }\bibfield  {title} {\enquote {\bibinfo
  {title} {Influence of water evaporation/absorption on the stability of
  glycerol-water marbles},}\ }\href {\doibase 10.1039/c9ra05728e} {\bibfield
  {journal} {\bibinfo  {journal} {Roy. Soc. Ch. Adv.}\ }\textbf {\bibinfo
  {volume} {9}},\ \bibinfo {pages} {34465--34471} (\bibinfo {year}
  {2019})}\BibitemShut {NoStop}%
\bibitem [{not({\natexlab{b}})}]{note}%
  \BibitemOpen
  \href@noop {} {} ({\natexlab{b}}),\ \bibinfo {note} {Note
  that the evolution of bubbles of different initial water glycerol mass ratio
  were followed over 6 hours for more than 320 bubbles, 24 hours for 36 of them
  and more than one year for 6 of them. For all these bubbles no naturally
  occurring rupture event has been observed after the first 4 hours of
  observation indicating that bubble, which last more than 4 hours will not
  break afterwards. This is consistent with figure 2d which shows that the time
  required to reach a stable state is less than 4 hours. Hence, we have chosen
  a 50\% longer time (6h) than the last observed rupture (4h) as a reference
  time to define “everlasting bubbles”.}\BibitemShut {Stop}%
\bibitem [{\citenamefont {Dollet}\ and\ \citenamefont
  {Boulogne}(2017)}]{convection_evaporating}%
  \BibitemOpen
  \bibfield  {author} {\bibinfo {author} {\bibfnamefont {Benjamin}\
  \bibnamefont {Dollet}}\ and\ \bibinfo {author} {\bibfnamefont
  {Fran\ifmmode\mbox{\c{c}}\else\c{c}\fi{}ois}\ \bibnamefont {Boulogne}},\
  }\bibfield  {title} {\enquote {\bibinfo {title} {Natural convection above
  circular disks of evaporating liquids},}\ }\href {\doibase
  10.1103/PhysRevFluids.2.053501} {\bibfield  {journal} {\bibinfo  {journal}
  {Phys. Rev. Fluids}\ }\textbf {\bibinfo {volume} {2}},\ \bibinfo {pages}
  {053501} (\bibinfo {year} {2017})}\BibitemShut {NoStop}%
\bibitem [{\citenamefont {Miguet}\ \emph {et~al.}(2020)\citenamefont {Miguet},
  \citenamefont {Pasquet}, \citenamefont {Rouyer}, \citenamefont {Fang},\ and\
  \citenamefont {Rio}}]{big_bubbles}%
  \BibitemOpen
  \bibfield  {author} {\bibinfo {author} {\bibfnamefont {J.}~\bibnamefont
  {Miguet}}, \bibinfo {author} {\bibfnamefont {M.}~\bibnamefont {Pasquet}},
  \bibinfo {author} {\bibfnamefont {F.}~\bibnamefont {Rouyer}}, \bibinfo
  {author} {\bibfnamefont {Y.}~\bibnamefont {Fang}}, \ and\ \bibinfo {author}
  {\bibfnamefont {E.}~\bibnamefont {Rio}},\ }\bibfield  {title} {\enquote
  {\bibinfo {title} {Stability of big surface bubbles: Impact of evaporation
  and bubble size},}\ }\href {\doibase 10.1039/c9sm01490j} {\bibfield
  {journal} {\bibinfo  {journal} {Soft Matter}\ }\textbf {\bibinfo {volume}
  {16}},\ \bibinfo {pages} {1082--1090} (\bibinfo {year} {2020})}\BibitemShut
  {NoStop}%
\bibitem [{\citenamefont {D'Errico}\ \emph {et~al.}(2004)\citenamefont
  {D'Errico}, \citenamefont {Ortona}, \citenamefont {Capuano},\ and\
  \citenamefont {Vitagliano}}]{diffusion_coefficient}%
  \BibitemOpen
  \bibfield  {author} {\bibinfo {author} {\bibfnamefont {G.}~\bibnamefont
  {D'Errico}}, \bibinfo {author} {\bibfnamefont {O.}~\bibnamefont {Ortona}},
  \bibinfo {author} {\bibfnamefont {F.}~\bibnamefont {Capuano}}, \ and\
  \bibinfo {author} {\bibfnamefont {V.}~\bibnamefont {Vitagliano}},\ }\bibfield
   {title} {\enquote {\bibinfo {title} {Diffusion coefficients for the binary
  system glycerol + water at 25 °c. a velocity correlation study},}\ }\href
  {\doibase 10.1021/je049917u} {\bibfield  {journal} {\bibinfo  {journal} {J.
  Chem. Eng. Data}\ }\textbf {\bibinfo {volume} {49}},\ \bibinfo {pages}
  {1665--1670} (\bibinfo {year} {2004})}\BibitemShut {NoStop}%
\bibitem [{\citenamefont {Association}(1963)}]{properties_glycerine}%
  \BibitemOpen
  \bibfield  {author} {\bibinfo {author} {\bibfnamefont {Glycerine~Producers'}\
  \bibnamefont {Association}},\ }\href
  {https://books.google.fr/books?id=XpeaGQAACAAJ} {\emph {\bibinfo {title}
  {Physical Properties of Glycerine and Its Solutions}}}\ (\bibinfo
  {publisher} {Glycerine Producers' Association},\ \bibinfo {year}
  {1963})\BibitemShut {NoStop}%
\end{thebibliography}
\end{document}